# Automated Detection of Algorithm Debt in Deep Learning Frameworks: An Empirical Study

Emmanuel Iko-Ojo Simon[1], Chirath Hettiarachchi[2], Alex Potanin[3], Hanna Suominen[4], Fatemeh Fard[5]

*School of Computing, College of Engineering, Computing & Cybernetics, Australian National University, Australia*[1,2,3,4]
*ANU College of Health and Medicine, School of Medicine and Psychology*[4]
*University of Turku, Faculty of Technology, Department of Computing, Turku, Finland*[4]
*Department of Computer Science, University of British Columbia, Kelowna, Canada*[5]
emmanuel.simon@anu.edu.au[1]

*Abstract*—Context: Previous studies demonstrate that Machine or Deep Learning (ML/DL) models can detect Technical Debt from source code comments called Self-Admitted Technical Debt (SATD). Despite the importance of ML/DL in software development, limited studies focus on automated detection for new SATD types: Algorithm Debt (AD). AD detection is important because it helps to identify TD early, facilitating research, learning, and preventing the accumulation of issues related to model degradation and lack of scalability.
Aim: Our goal is to improve AD detection performance of various ML/DL models.
Method: We will perform empirical studies using approaches: TF-IDF, Count Vectorizer, Hash Vectorizer, and AD-indicative words to identify features that improve AD detection, using ML/DL classifiers with different data featurisations. We will use an existing dataset curated from seven DL frameworks where comments were manually classified as AD, Compatibility, Defect, Design, Documentation, Requirement, and Test Debt. We will explore various word embedding methods to further enrich features for ML models. These embeddings will be from models founded in DL such as ROBERTA, ALBERTv2, and large language models (LLMs): INSTRUCTOR and VOYAGE AI. We will enrich the dataset by incorporating AD-related terms, then train various ML/DL classifiers: Support Vector Machine, Logistic Regression, Random Forest, ROBERTA, and ALBERTv2.

## I. INTRODUCTION

Deep Learning (DL) and conventional Machine Learning (ML) technologies have advanced in their applications in various sectors, including healthcare, tourism, finance, business, and education [19, 50, 48]. These advancements include the emergence of applications such as GPT-4o, voice assistants, and computer vision applications like self-driving cars [7].

Just like traditional software systems, ML/DL systems are not immune to Technical Debt (TD) [30]. Traditional software refers to software systems that follow standard programming paradigms and do not incorporate any ML/DL algorithms or components, while ML/DL systems are systems that integrate ML/DL models or algorithms with other non-ML components [7]. TD was introduced as a metaphor by Cunningham [11], referring to the cost of having short-term benefits during software development at the expense of long-term future maintenance. *TD symbolises the tacit compromise between delivering fast and producing high-quality code* [37].

Due to time pressure, market competition, and cost reduction, DL framework developers are often faced with the choice of shorter time completion and better software quality, leading to compromised decisions and increased burden in future maintenance [25]. In a work by Liu et al. [25], they uncovered two new TD types prevalent in DL frameworks — Algorithm Debt (AD) and Compatibility Debt (CD). CD refers to debt related to a project's immature dependencies on other projects, which cannot supply all qualified services, and the current implementation is a temporary workaround. In this work, we focus on AD, which refers *to sub-optimal implementations of algorithm logic*.

AD in ML/DL systems poses unique challenges, negatively impacting the software as it evolves with time. Neglecting AD could risk performance issues, model degradation, inconsistent results, and hinders scalability [42, 47, 4, 53]. Focusing on AD allows a deeper exploration of its distinct characteristics compared to CD [25] and other ML/DL-related TD types [42]. This focused approach lays the groundwork for understanding the nuances of AD, considering its nascent nature, and could encourage further research into CD and other TD types in ML/DL systems. Thus, not focusing on CD or other TD types in this study does not diminish their importance.

Previous research on traditional software have demonstrated that ML/DL classifiers can automatically detect TD from source code comments that highlight TD, often referred to as Self-Admitted Technical Debt (SATD) [29, 16, 43]. The automatic detection of AD is important because early identification of AD could help to mitigate AD-related issues such as model degradation, inconsistent results, and lack of scalability [42, 47, 4, 53]. This will improve the overall performance, reliability, and maintainability of ML/DL systems, ensuring they can adapt to evolving requirements.

The comment illustrated in Listing I, originating from

Listing 1. Example of Algorithm Debt on Line 1 in an Open-Source System.

```
// TODO(Yangqing): Is there a faster way to do pooling in the channel-first case?
template <typename Dtype>
void PoolingLayer<Dtype>::Forward_cpu(const vector<Blob<Dtype>*>& bottom,
    const vector<Blob<Dtype>*>& top) {
  const Dtype* bottom_data = bottom[0]->cpu_data();
  Dtype* top_data = top[0]->mutable_cpu_data();
  const int top_count = top[0]->count();
}
```

an open-source system called caffe[1] highlights a potential inefficiency in the pooling operation for the channel-first case. The developer has acknowledged the need for a more efficient algorithm but has not implemented it yet. This deferred optimisation represents AD as it introduces TD which can degrade performance over time potentially leading to slower execution times. Inefficient pooling operations can become a bottleneck, especially when dealing with large-scale data, thereby limiting the scalability of the system [4]. Addressing such comments early could reduce AD and enhance the system's overall performance.

While prior works explored SATD in traditional software without any ML/DL components, SATD in ML/DL is different. For example, the SATD in Listing I on inefficiency in the pooling operation cannot be found in traditional software and is specific to ML/DL systems. Studies show that the median percentage of SATD is twice as prevalent in ML/DL systems compared to traditional software and appears earlier in the ML/DL development life cycle [25, 7]. Furthermore, the dependence on data and the lack of developer awareness about the intricacies of ML/DL models contribute significantly to SATD in these systems [31]. In traditional software, SATD usually manifests as code debt [5], while in ML/DL projects it often takes the form of requirements debt [25].

The automated detection of SATD in previous studies focused on traditional programming with limited studies focused on the DL domain. Furthermore, the classifier proposed for detecting SATD by Sharma et al. [43] reported an F1 score of 31% for AD when using the ROBERTA model, suggesting room for improvement. Exploring various ML/DL classifiers will help identify the most effective approach for AD detection. Unlike previous works with a limited set of classifiers, our evaluation involves different classifiers and featurisations, each selected for its potential to address AD.

In this report, we describe our aim to *improve AD detection performance* in DL frameworks through empirical studies using different approaches: Term Frequency- Inverse Document Frequency (TF-IDF), Count Vectorizer, Hash Vectorizer, and AD-indicative words to identify features that will improve AD detection, using ML/DL classifiers with different data featurisation. We will train the ML/DL classifiers on a dataset curated by Liu et al. [25] from seven open-source DL frameworks. The frameworks include: TensorFlow, Keras, DL4J, Caffe, PyTorch, MXNet, and Microsoft Cognitive Toolkit. We selected this dataset because it was the first to uncover AD, thus minimising the risk of misclassifying AD instances.

The ML classifiers we will use are the Support Vector Machine (SVM) [17], Logistic Regression (LR) [15], and Random Forest (RF) [8]. For DL models, we will use ROBERTA [28], and ALBERTv2 [20] and their embeddings. We selected these ML/DL models because they were used in previous SATD detection tasks [23, 39, 43].

Acknowledging the advancements in large language models (LLMs) based on DL, we will explore the INSTRUCTOR [44] and VOYAGE AI[2] models to obtain the embeddings of SATD comments. These LLMs have achieved the state-of-the-art performance in SE tasks such as automatic code generation, bug detection, and natural language tasks [21, 32, 27, 40].

As part of our research, we plan to enrich the SATD dataset by incorporating additional features related to AD to improve the robustness of our models. This is to provide more context and allow the classifiers to better capture the nuances of AD, which could lead to improved detection performance.

Since AD is still novel, the findings from this research could lead to computer-assisted analysis aiding in disentangling and understanding what AD means in the specific context of ML/DL systems and the timeline of engineering them.

## II. RELATED WORKS ON TD

### A. Studies on TD

Cunningham [11] introduced the concept of TD as "quick and dirty" work in software design and results in code that is "not quite-right". However, TD is an abstract concept, since making tradeoffs between optimal software and meeting project deadlines is not quantitatively measurable [7]. Hence, Potdar and Shihab [35] introduced SATD, which was deemed to be a more visible measure of TD. They examined SATD in large open-source projects and found a high prevalence of up to 31% in files. Building on this, Maldonado et al. [29] studied the evolution of traditional code and revealed that SATD lasts between 18 and 172 days in a system. Bavota and Russo [5] performed a large-scale study of SATD across different software projects, contributing towards developing a taxonomy of SATD for traditional software.

Several other studies have been done on TD [1, 33]. Lenarduzzi et al. [22] performed a systematic literature review on TD and proposed different approaches for prioritising TD. Similarly, Xiao et al. [51], conducted a large-scale study on SATD comments to investigate the prevalence of SATD clones. They observed that SATD clones are a more prevalent phenomenon in build systems than in source code.

---
[1]https://github.com/BVLC/caffe/blob/master/src/caffe/layers/pooling_layer.cpp

[2]https://www.voyageai.com/

*B. SATD Detection in Traditional Software Engineering*

Since the first study on SATD [35], different approaches have been used to automatically identify SATD in traditional software engineering. da Silva Maldonado et al. [12] used the NLP Max Entropy to identify SATD from ten open-source projects. They classified the SATD into 5 types and obtained an average F1-score of 62% for design and 43% for requirements TD. Wattanakriengkrai et al. [49] introduced an approach to SATD identification using N-gram IDF to identify comments as design, requirements, or non-SATD. They used the RF model and obtained an average F1-score of 64%.

In another work by Sharma et al. [43], they studied the capabilities of different classifiers – ME, SVM and PLMs to classify SATD in R. Their results showed that the macro-averaged F1-score of the various classifiers was between 42% and 56%. Recently, Li et al. [23] in their work proposed an approach for automated identification of SATD from source code comments, commit messages, pull requests, and issue tracking systems using LR, SVM, RF and Text-CNN. They achieved an average F1-score of 61%.

Sabbah and Hanani [39] explored different models: SVM, RF, Naive Bayes, and CNN to identify SATD from code comments or commits. They achieved an average F1-score of 82% and 84% for RF and CNN respectively. In another work, Pinna et al. [34], investigated SATD in Blockchain, using the the Stanford classifier. They obtained an F1-score of 48% and 54% for design and requirements debt.

*C. TD in ML/DL*

Sculley et al. [42] first conducted pioneering research into TD risk factors that are specific to ML systems. They identified key factors that contribute to TD accumulation and the importance of managing TD in ML systems. Liu et al. [25] examined the prevalence of TD in DL frameworks and found that design, defect, and documentation debt are the most prevalent TD in DL frameworks. Additionally, this work uncovered two new TD types in DL frameworks: AD and CD. As a follow-up, Liu et al. [26] in their work investigated the introduction and removal of SATD in DL frameworks and found that design debt is introduced the most along the development process.

Bhatia et al. [7] studied the occurrence of SATD in ML code by manually analysing ML projects across five domains. They found that ML pipeline components for data preprocessing are more susceptible to TD than other components.

*D. Novelty*

The existing studies on the automatic detection of SATD were focused mostly on traditional software which is different from the DL domain [3]. Also, the classifier by Sharma et al. [43] reported an F1 score of 31% for AD, suggesting room for improvement. In our work, we seek to improve the detection of SATD in DL systems using ML/DL techniques with an emphasis on AD. We plan on:

- Focusing on the DL domain, which has been relatively under-explored in the context of automated detection of SATD using ML/DL models.
- Targeting AD, which has not been extensively studied in existing literature.
- Introducing a novel approach for SATD detection that enriches the SATD dataset with additional context, moving beyond the focus on code comments alone.

III. RESEARCH QUESTION (RQ)

The aim of this research is to improve AD detection performance of various ML/DL models. We will perform empirical studies with different features (e.g., TF-IDF, count vectorizer, Hash Vectorizer) to identify which features will improve the performance of ML/DL classifiers in AD detection. To achieve this, we formulated our goal using GQM [9]:

**Purpose:** Improve
**Issue:** AD detection performance
**Object:** of various ML/DL models in DL frameworks
**Viewpoint:** from SATD.

Therefore, we refined our goal to two RQs:

RQ1: Which feature extraction methods enhance the performance of ML/DL classifiers in detecting AD? Identifying the most effective features is essential for improving AD detection accuracy. We will conduct empirical studies with various feature extraction techniques to evaluate their impact on the performance of different ML/DL models

RQ2: How do different ML/DL models perform in detecting AD? Comparing the performance of ML/DL models with various features could help identify the best model for AD detection. We will assess the capabilities of different ML/DL classifiers using various feature extraction methods.

IV. DATASET

We will utilise the dataset curated by Liu et al. [25], which contains SATD comments, to train and evaluate our classifiers. Using this dataset reduces the chances of personal bias in labeling new data since this research uncovered AD. Also, this dataset had a high inter-rater agreement (Cohen's Kappa coefficient of +85%), ensuring the reliability of the data.

The dataset contains comments indicating SATD in seven open-source DL frameworks hosted on GitHub. The frameworks include TensorFlow, Keras, DL4J, Caffe, PyTorch, MXNet and Microsoft Cognitive Toolkit (CNTK). The authors manually classified all SATD comments in the latest stable version of these frameworks into one of seven categories: Algorithm, Compatibility, Defect, Design, Documentation, Requirements, and Test. The dataset has a total of 15,719 SATD instances, specifically with with AD having a total of 935. The percentage distribution is shown in Table I.

Focusing on data from DL frameworks ensures that the ML/DL classifiers will be trained with a DL domain-specific data. More so, the dataset contains terms unique to DL frameworks such as **RNN, LSTM, CUDNN, BATCHNORM**. We believe that this focused approach will take the unique characteristics of DL frameworks into perspective.

TABLE I
PERCENTAGE DISTRIBUTION OF TD TYPES. T.FLOW STANDS FOR
TENSORFLOW AND P.TOR STANDS FOR PYTORCH. ALSO, AD –
ALGORITHM, CD – COMPATIBILITY, DD – DEFECT, DsD – DESIGN, DoD
– DOCUMENTATION, RD – REQUIREMENT, AND TD –TEST DEBT

|     | T.Flow | Keras | Caffe | P.Tor | MXNet | CNTK  | DL4J  |
| --- | ------ | ----- | ----- | ----- | ----- | ----- | ----- |
| AD  | 7.09   | 31.48 | 10.00 | 10.45 | 10.16 | 10.53 | 13.92 |
| CD  | 3.78   | 35.18 | 11.87 | 7.37  | 2.96  | 3.95  | 0.21  |
| DD  | 5.53   | 1.85  | 5.00  | 4.20  | 8.47  | 7.48  | 8.22  |
| DsD | 57.03  | 24.07 | 48.75 | 59.73 | 63.13 | 65.27 | 55.90 |
| DoD | 1.27   | 0.00  | 15.62 | 0.81  | 0.42  | 0.83  | 0.42  |
| RD  | 17.56  | 7.40  | 5.62  | 12.80 | 11.01 | 7.41  | 20.67 |
| TD  | 7.70   | 0.00  | 3.12  | 4.61  | 3.81  | 4.50  | 0.63  |

## V. EXECUTION PLAN

### A. Data Cleaning

We will follow a standard process used in previous works for data cleaning [29, 34]. This will ensure the quality of data and consistency used for the training of the classifiers. We will remove stop words that frequently appear since they do not add any important meaning to the dataset [16] using *nltk*[3]. Leveraging results from Huang et al. [16], we will remove punctuation from the comments, with the exception of '!' and '?'. This practice is important because it has been demonstrated that retaining these two specific punctuation marks will help to improve SATD detection [34]. Subsequently, for data uniformity and to remove the encoding differences, we will convert all tokens to lowercase [16] using `nltk`. We will also convert numbers to their respective numerals using `wordToNum`. This will ensure consistency in the use of terms across the training data.

### B. ML/DL Classifiers and Embedding Models

Firstly, we will explore the ML classifiers: SVM [17], LR [15], and RF [8] because they have been explored in previous SATD classification [43, 23] and have specific features beneficial for AD detection. **SVM** is effective in high-dimensional spaces and has been shown to perform well in text classification tasks, with the ability to handle large feature sets, making it a suitable choice for AD detection [17]. **LR** classifier is efficient for multi-class classification problems and performs well in scenarios with a large number of features, making it a good fit for AD detection [15]. **RF** is an ensemble learning method that combines multiple decision trees to improve detection performance. It is robust to overfitting and can handle the complex relationships in the data, which is beneficial for detecting AD [8]. We will train these classifiers using a supervised learning approach based on previous studies [43, 24].

The creation of embeddings that capture patterns within data is a major breakthrough in DL. Learned embeddings from DL models, such as those generated by transformer-based architectures like BERT [13], can process natural language. These embeddings offer high-dimensional data representations that retain semantic information, enabling the training of efficient ML models. We will use the embeddings generated by DL models, **ROBERTA**[4] and **ALBERTv2**[5], to train several ML classifiers. Similar to previous studies, we will also train these DL models to detect SATD [43, 36]. These DL models have been trained on a large dataset and can be fine-tuned for a downstream task.

**ROBERTA** is a variant of BERT that has been fine-tuned to perform exceptionally well in various NLP tasks. Its ability to handle contextual understanding in a text makes it highly suitable for detecting SATD [28]. **ALBERTv2** is a lighter and faster variant of BERT that reduces memory consumption while maintaining performance. Previous studies have demonstrated its effectiveness in the classification of SATD and non-SATD, making it a logical choice for our experiments [20].

We also will explore state-of-the-art LLMs approaches founded in DL used for text classification tasks: **INSTRUCTOR**[6] [44] and **VOYAGE AI**[7] to embed the comments. The **INSTRUCTOR** [44] model was designed to provide high-quality embeddings and is effective in text classification tasks; hence our choice to explore it for detecting AD. **VOYAGE AI** [46] is another LLM founded on DL known for its performance in text classification tasks. We will use this model to extract embeddings of comments. After we embed the comments, we will then apply ML models.

*Baselines:* To mitigate the risk of bias and subjectivity that could arise if we were to create our own baseline, we will compare our results with previously established works on SATD detection. In addition, we will use a random classifier to ensure that our models are not predicting random values.

**Random Classifier:** A random classifier provides a fundamental baseline by assigning labels randomly, without considering any features or patterns in the data. This helps in establishing a lower bound of performance and allows us to gauge how much better our method performs compared to random chance. The next baseline is the approach by **Sharma et al. [43]** as it is the only existing work that performed SATD classification including AD. The ROBERTA model – the best in their study, had an F1-Score of 31%, making it a relevant baseline. Using this as a baseline enables us to directly evaluate how well our method performs against the state-of-the-art in detecting AD. The work by **Sabbah and Hanani [39]** is among the latest in SATD detection, employing a Convolutional Neural Network (CNN) for multi-classification of SATD types. The use of CNNs for text classification tasks is well-established and this makes it a suitable baseline for evaluating the effectiveness of our approach. The approach by **Li et al. [24]**, known as MT-Text-CNN, is a recent work in SATD detection that leverages multitasking learning tech-niques. This method uses a convolutional neural network to identify multiple SATD types. Using this model as a baseline allows us to compare our model to one of the most recent methods in the field.

---

[3] https://www.nltk.org/
[4] https://huggingface.co/docs/transformers/model_doc/roberta
[5] https://huggingface.co/tftransformers/albert-base-v2
[6] https://instructor-embedding.github.io/
[7] https://www.voyageai.com/

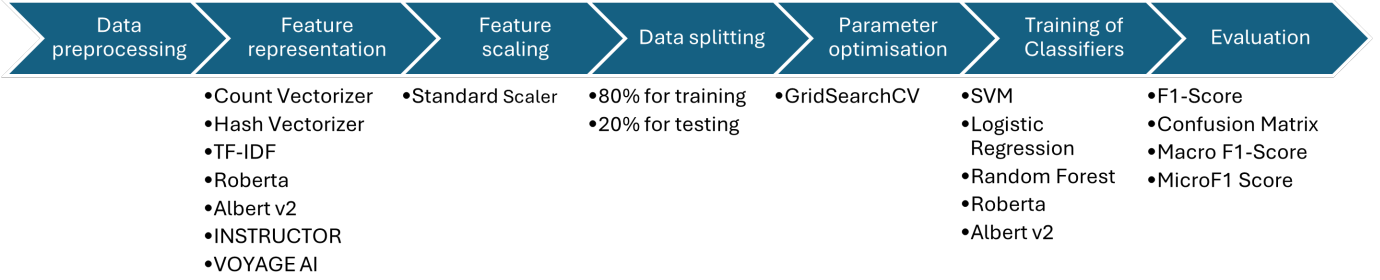

Fig. 1. The steps used in training the classifiers

*C. Model Training, Tuning, and Testing*

The steps that we will follow to train the classifiers are shown in Figure 1. For the ML classifiers, we will use the data from Section V-A, by transforming the SATD comments into feature representations by using different word embedding techniques: the TF-IDF [39], CountVectorizer, and HashVectorizer. We will also use the embeddings from DL models such as ROBERTA [28], ALBERTv2 [20], as well as DL LLMs INSTRUCTOR [44] and VOYAGE AI[8] to embed the comments, where each comment will be represented as a fixed length embedding before training lightweight ML classifiers. After applying the feature representation techniques above, we will transform the textual data into a structured matrix of token counts, with each row representing a distinct document and each column signifying a unique word within the corpus.

We will then standardise the data using the *Standard Scaler*, to ensure uniformity across all features in the dataset. The Standard Scaler standardises features by removing the mean and scaling to unit variance. The standard score of a sample $x$ is calculated as $z = (x − \mu)/\sigma$, where $\mu$ is the mean of the training samples, and $\sigma$ is the standard deviation of the training samples. We note that this feature engineering will not be necessary for the DL models, so we will only apply it to the ML models. Similarly, we will not use embeddings for the DL models, since they already incorporate embedding layers as part of their architecture [14].

For the traditional ML classifiers, we will conduct parameter optimisation using Grid Search CV [6]. Grid search is a hyperparameter tuning approach that evaluates each possible parameter combination in the search space. We will use the Grid Search CV to enable us to identify the optimal settings for our dataset appropriate for our text classification. For SVM, we will experiment with different kernels such as *RBF, linear, and polynomial*. For LR, we will explore parameters such as *C*(inverse of regularization strength) and penalty (type of regularization, such as *L1 or L2*). Likewise, for the RF classifier, we will experiment with hyperparameters including *Criterion, maximum sample size, and minimum leaf size*. We will not use GridSearch CV nor run multiple experiments for the DL models. This is because DL models have a vast number of hyperparameters, and exhaustive hyperparameter tuning using GridSearchCV can become computationally expensive [18]. Consequently, the benefits of running multiple experiments with varying hyperparameters may not outweigh the associated computational costs and time.

[8]https://www.voyageai.com/

We will split the resulting dataset from Section V-A into training and testing sets, in the ratio of 80:20, following the methodology used in a similar task [43]. This will ensure that the data we will use for testing remains unseen during training, thereby mitigating the chances of overfitting. For the training set, we will implement stratified *k*-fold cross-validation [52] with $k = 10$. This method was used in previous studies [23, 10] and this is to ensure robust evaluation. Specifically, we will split the 80% training data into 10 equal parts, maintaining the class distribution in each fold. During each iteration of the 10-fold cross-validation, we will use nine folds for training the model and use the remaining fold as a validation set for tuning hyperparameters. This process will be repeated 10 times, each time with a different fold serving as the validation set. We will use the 20% test set only for the final evaluation of the model after hyperparameter tuning and training are complete. This will ensure that the evaluation metrics are not skewed due to class imbalances in the dataset and that the models are properly tuned before the final evaluation.

We will use metrics such as precision, recall, F1-score, and Average F1-scores to evaluate our model consistent with studies on SATD detection [38, 41, 43, 23]. For instance, Sharma et al. [43] used precision, recall, and micro-averaged F1-scores to evaluate their SATD classifiers. Similarly, previous studies to classify Design and Requirements Debt used precision, recall, and F1-scores in their work [38, 41, 23]. By using these established metrics, we ensure the comparability and robustness of our evaluation.

*D. Data Enrichment*

Taking inspiration from the work of Suominen et al. [45] in a previous text classification task, we will enrich our data. This is by adding the textual definitions of each TD type found in our dataset. These definitions will be from established works on TD [2, 25]. The aim of adding the extra data is not because the data is insufficient but to add context for the ML/DL classifiers. We believe that this will potentially improve their ability to detect SATD instances more accurately. For example, for AD, we will add: *AD is a TD type prevalent across multiple domains, resulting from the sub-optimal implementation of an algorithm or logic. It is characterised by its negative impact on system performance such as model degradation, lack of scalability and other quality attributes as the system evolves.*

Also, in the feature engineering for our approach, we plan to incorporate AD-specific terms tailored to extract AD-related keywords, further capturing the nuances of AD in DL frameworks, hence making our approach specific to AD in DL.

*E. Statistical Significance Tests*

To validate the performance of our models, we will incorporate statistical significance tests in our experiments to test if the results are significant. This analysis will be used to verify if the differences in precision, recall, and F1-score are statistically significant by calculating the p-values[9] for these metrics. The results will be evaluated at a significance level of 0.05, implying that p-values less than this threshold will confirm statistically significant improvements in model performance. Additionally, we will assess how each TD type's total sample count and unique number of words related to the F1-scores by computing the p-values.

## VI. THREATS TO VALIDITY

We will consider several threats to the validity of this research. On **internal validity**, to mitigate potential researcher bias from the manual classification process, we will use an already classified dataset from previous research [25] to train our ML/DL models. The authors reported a high level of agreement between the two annotators (Cohen's kappa of +85%), giving us confidence in the dataset. To mitigate potential bias in the classification results, we will use stratified 10-fold cross-validation, ensuring each sample is used in the test set once while preserving class distribution in the train and test sets.

The **External validity** of our study is limited by the scope of our research, which will focus exclusively on DL frameworks. Therefore, the generalisability of our findings on AD is specific to the domain of DL and may not be extended to include traditional software engineering. Nevertheless, the relevance of the study remains unaffected.

Threats to **construct validity** refer to the suitability of our evaluation metrics. We will use precision, recall, and F1-score, consistent with past studies evaluating automated SATD detection approaches [43, 38, 41, 23, 39].

---

[9]The p-value measures the probability of obtaining results at least as extreme as the observed ones, assuming the null hypothesis is true. A p-value less than 0.05 indicates statistically significant results.

## VII. DATA AVAILABILITY

To ensure the replicability of this work and to adhere to ACM's open access policy, we will create an online repository in GitHub[10], which will include all the scripts and the dataset.

## VIII. CONCLUSION

The objective of this study plan is to improve AD detection performance of various ML/DL models. We will perform empirical studies with different features to identify which features will improve the performance of AD detection using ML/DL classifiers with different data featurisations.

We will employ qualitative approaches to address the goal of our investigation and, based on the findings we will be able to determine the best classifier for AD detection in DL systems, for researchers to explore further. Additionally, this study will make a valuable contribution to the existing body of knowledge on empirical methods, as it will be the pioneering work that will enhance the SATD dataset with a collection of AD indicators for SATD detection.

## ACKNOWLEDGEMENTS

I would like to express my gratitude to the Australian National University for providing the funding through the ANU PhD scholarship within the ANU Research School of Computing, College of Engineering, Computing and Cybernetics.

---

[10]https://github.com/